# Lévy Noise Induced Switch in the Gene Transcriptional Regulatory System

Yong Xu[†], Jing Feng, JuanJuan Li, Huiqing Zhang

*Department of Applied Mathematics, Northwestern Polytechnical University, Xi'an 710072, China*

**Abstract:** Gene transcriptional regulatory is an inherently noisy process. In this paper, the study of fluctuations in a gene transcriptional regulatory system is extended to the case of Lévy noise, a kind of non-Gaussian noises which can describe unpredictable jump changes of the random environment. The stationary probability density is given to explore the key roles of Lévy noise in the gene regulatory networks. The results demonstrate that the parameters of Lévy noise, including noise intensity, stability index and skewness parameter can induce switches between distinct gene-expression states. A further concern is the switching time (from the high concentration state to the low concentration one or from the low concentration state to the high concentration one), which is a random variable and often referred to as the mean first passage time. The effects of Lévy noise in expression time and degradation time is studied by computing the mean first passage time in two directions and a number of different peculiarities of non-Gaussian Lévy noise compared with Gaussian noise are observed.

## I. Introduction

Regulation of gene expression by signals from outside and within the cell is very important for cells to control fundamental functions, and nonlinear ordinary differential equations are applied to model gene regulation which have been widely considered to gain insight into the underlying processes of living systems[1,2,3]. The previous work regarded the production of gene product as a deterministic process with no random components, while random fluctuation systems can behave quite differently from deterministic ones in biology systems[4,5,6]. It is well documented that random noise have been observed with critical roles in clinical and experimental studies and then deterministic models should be augmented with noise terms

[†] Corresponding author. TEL/FAX: 86 29 8843 1657, E-mail: hsux3@nwpu.edu.cn

[7,8,9,10]. In [7], the roles of the additive noise and multiplicative noise were considered in engineered gene networks. Based on the model in [8], Liu and Jia considered the effects of fluctuations in the degradation and the synthesis reaction rate of the transcription factor, including the case of uncorrelated and correlated noises [9]. In [10], the authors extended the fluctuations in gene regulatory networks to the case of Gaussian colored noise.

The above investigations illustrated that random noise was a key component in gene regulation, and furthermore long jumps are always associated with a complex structure of the environment which could not be described as the Gaussian noise. So a kind of non-Gaussian noise needs to be considered to describe unpredictable jump changes of the random environment, and Lévy fights, a stochastic processe characterized by the occurrence of extremely long jumps, can solve this problem. The length of these jumps is distributed as a Lévy stable statistics which exhibit long tails and make the moments divergent. So far this kind of noise is frequently encountered in nature and it has been detected in biology, for example, in the behavioral patterns of certain species such as albatrosses [11] or zooplankton [12] and regulation of gene expression [13]. Gene expression is regulated through the interaction between transcription factors and their respective target sites. The cell achieves this regulation with a relatively small number of transcription factors searching massive lengths of DNA. When binding to a specified site of the genome, they can either activate or repress the transcription of the related gene. However, on searching for binding sites, Lévy search is regarded as an optimal search mechanism in contrast to locally over sampling Gaussian search [13,14,15].

Nowadays literatures evaluating the roles of Lévy noise in the regulation of gene expression are still lacking and we aim to fill up this gap. To provide specific evidence for the effects of Lévy noise, it will be incorporated into the master equation of a gene transcriptional regulatory system which will be simulated via Monte Carlo method. Bistability is often encountered in deterministic gene expression models, defined by the existence of two stable states which represent two levels of gene expressions, the high one referred to as “on”, and the low one referred to as “off” [16]. This property always implies the possibility of switch-like behavior: the system will remain near one of its stable states until a sufficiently large perturbation drives it into the vicinity of another stable state. Therefore we demonstrate the influences of Lévy noise through the construction of a protein switch and consider whether protein production can be turned “on” and “off” by using pulses.

We are also interested in the expected time that the system takes to switch from the low concentration to the high concentration and vice versa. The time is a random variable and is often referred to as the mean first passage time (MFPT), which is defined as the averaging time for a particle starting from a potential barrier to another one [17]. In [18], the authors presented a set of stochastic models and computed the MFPT of the spontaneous transitions caused by the stochasticity of the switches, and Liu and Jia [9] investigated the MFPT of a gene transcriptional regulatory system with the inner and outside Gaussian noise. Here, we calculate the MFPT for the gene switches and examine their behaviors under variations of several parameters of Lévy noise to explain the implications in the sense of gene regulation.

## II. Genetic Regulation Model

This section is to introduce the deterministic genetic regulation model. We first present a brief overview of $\lambda$-phage, a virus which infects the bacterium E.coli. As Fig.1 (a) shows, the key section of the phage DNA, the right operator region ($O_R$) is situated between two phage-encoded regulatory proteins: $\lambda$-repressor gene (also named *cI*) and *cro*, and the promoter $P_{RM}$ controls expression of *cI* and the $P_R$ promoter for the expression of *cro*. In Fig.1(b), the promoter region $O_R$ contains three tandem DNA binding sites named as $O_R1$, $O_R2$, and $O_R3$.

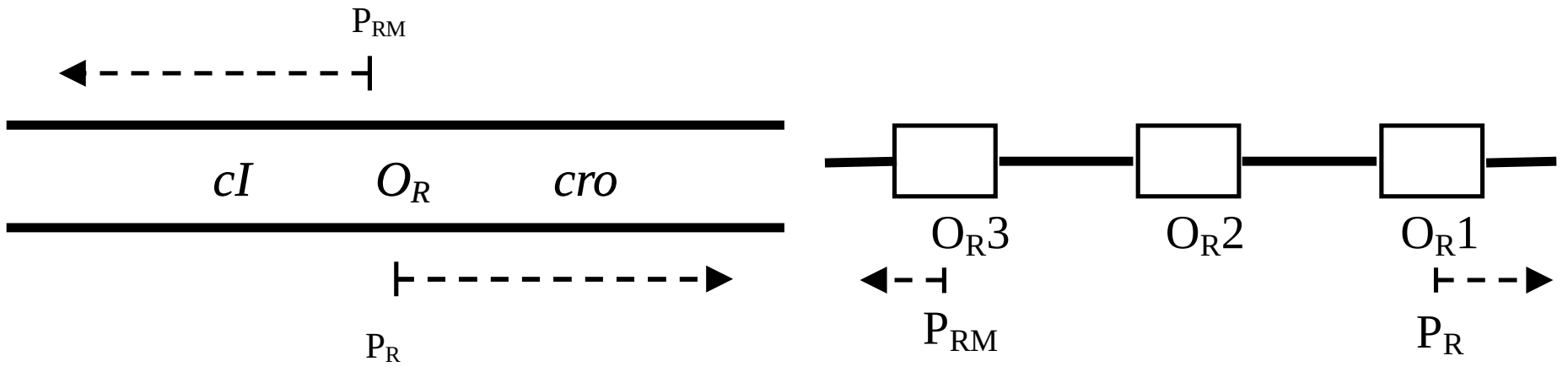

Fig.1. (a)The right operator region $O_R$ of $\lambda$ bacteriophage; (b) Three operator sites of $O_R$.

In [7], authors envisioned the system as a plasmid consisting of the $P_{RM}$ operator region and gene *cI* and only considered the regulation of $\lambda$-repressor (CI) expression. In this case, binding reactions can be given as follows: the gene *cI* expresses repressor (CI) which dimerizes in turn. Repressor dimmers then bind to the operator sites and regulate the transcription of repressor gene *cI*. Binding at $O_R1$ and $O_R2$ in the operator ($O_R$) activates transcription of its own gene by attracting RNA polymerase to the promoter ($P_{RM}$) of the repressor gene (*cI*). However, bounding to the site $O_R3$ represses transcription by excluding RNA polymerase from its own promoter ($P_{RM}$). A kinetic model according to these reactions can be written as [7]:

$$\dot{x} = \frac{k(2x^2 + 50x^4)}{25 + 29x^2 + 52x^4 + 4x^6} - rx + 1. \qquad (1)$$

Here, $x$ is the concentration of $\lambda$-repressor (CI). $k$ represents the increment of the transcription rate above the basal rate by the binding actions and $r$ stands for the degradation rate.

Then, bifurcation analysis is performed to study how the dynamic behaviors change versus the parameters $k$ and $r$. Fig.2 shows the bifurcation diagram for the system (1) as a function of $k$ and $r$. Note that, for values of $k$ and $r$ in the tinted region, there are three possible steady-state concentrations. Fig.3 is the phase graph for the system (1) in this case. It can be seen that the states $x_1$ and $x_3$ are stable because concentrations of CI near $x_1$ and $x_3$ will remain nearby despite slightly perturbation. The middle state $x_2$ is unstable, i.e. any small perturbations can drive the protein concentration towards $x_1$ or $x_3$. In the gap region, the system (1) has a single stable fixed point.

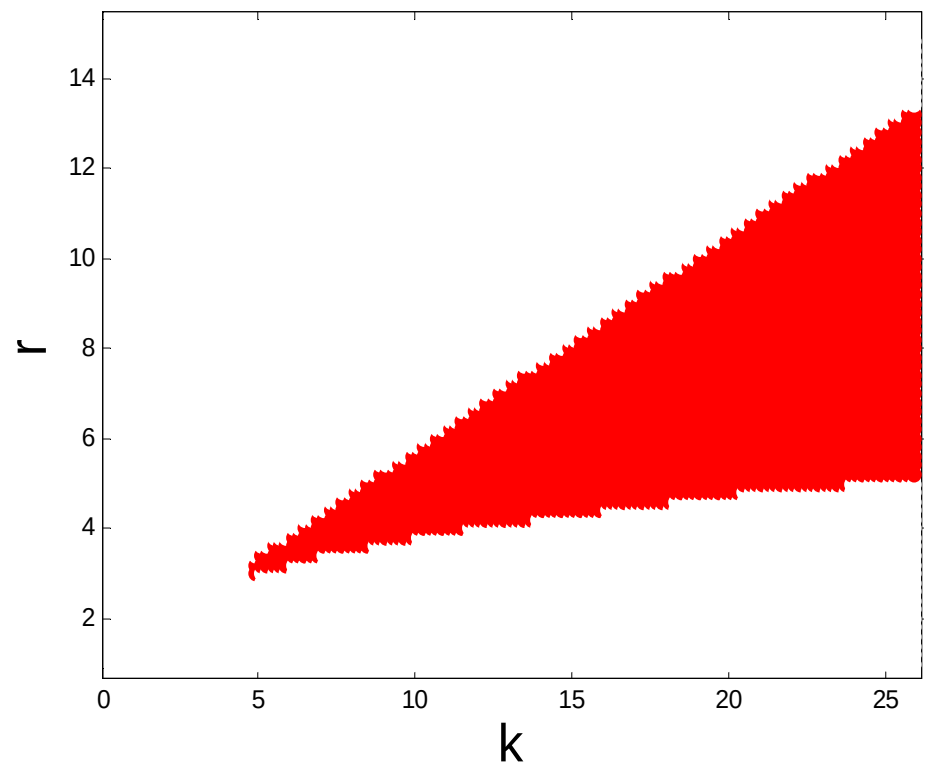

Fig.2. Deterministic bifurcation diagram for the system (1)

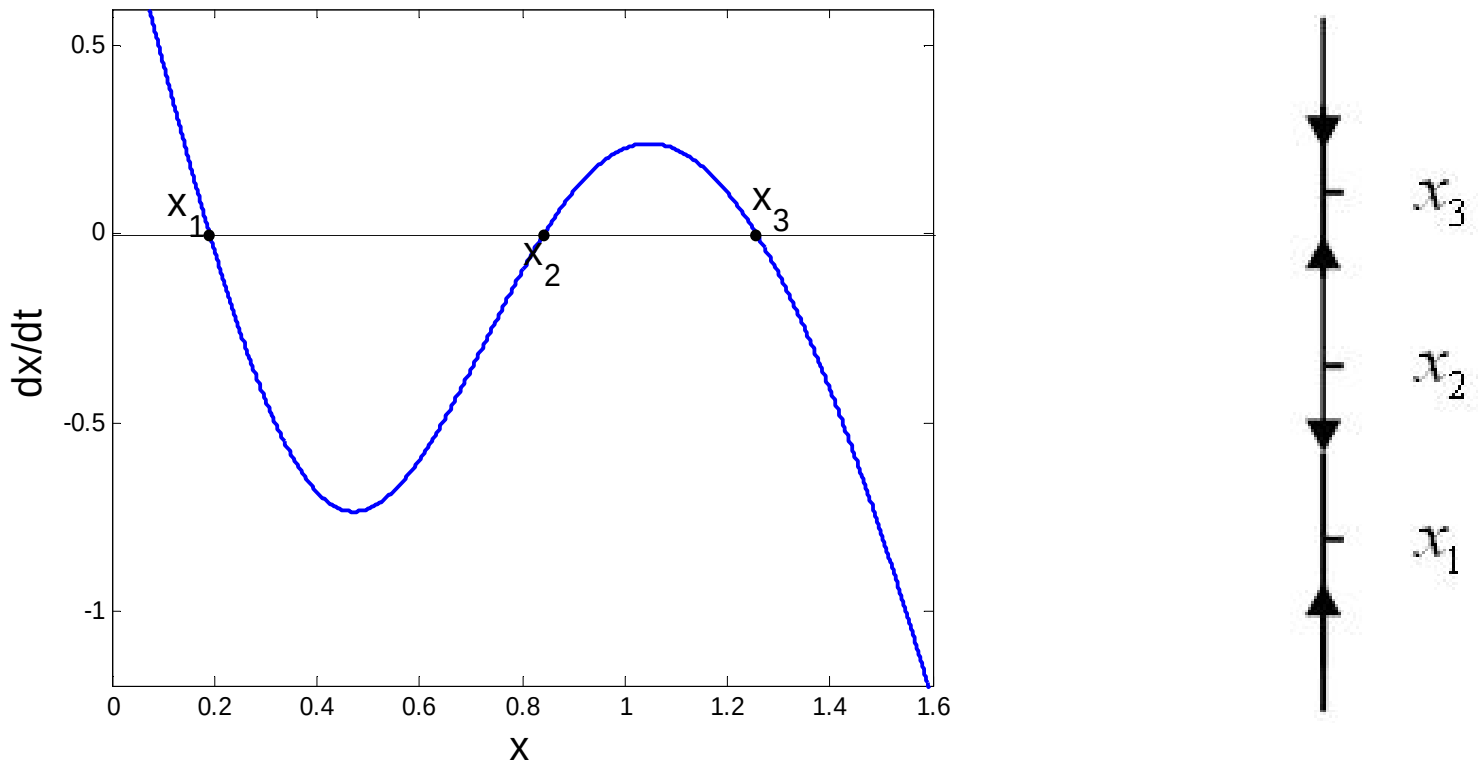

Fig.3. Phase diagram for the red region

## III. Lévy noise induced switch

In the deterministic case, when the system is bistability, for any given initial conditions, the system's trajectories will converge to one or the other of the equilibria and stay there for the future time. However, in response to environmental stimulation, the system can switch between two distinct steady states, a low-expression state where the protein is present in a small population and a high-expression state where the protein is present in a large population. Therefore, in this section, we focus on the values of $k$ and $r$ leading to bistability of the system and consider how Lévy noise affects the switch behaviors.

Here, we employ the stationary probability densities (SPDs) of the system driven by Lévy noise, which are able to characterize the contributions of Lévy noise to the behavior of the system [19]. We fix $k = 10, r = 5.5$ to exhibit bistability and then the potential function $U(x)$ of the deterministic system (1) is shown in Fig.4, whose local minima represent stable steady states of the concentration.

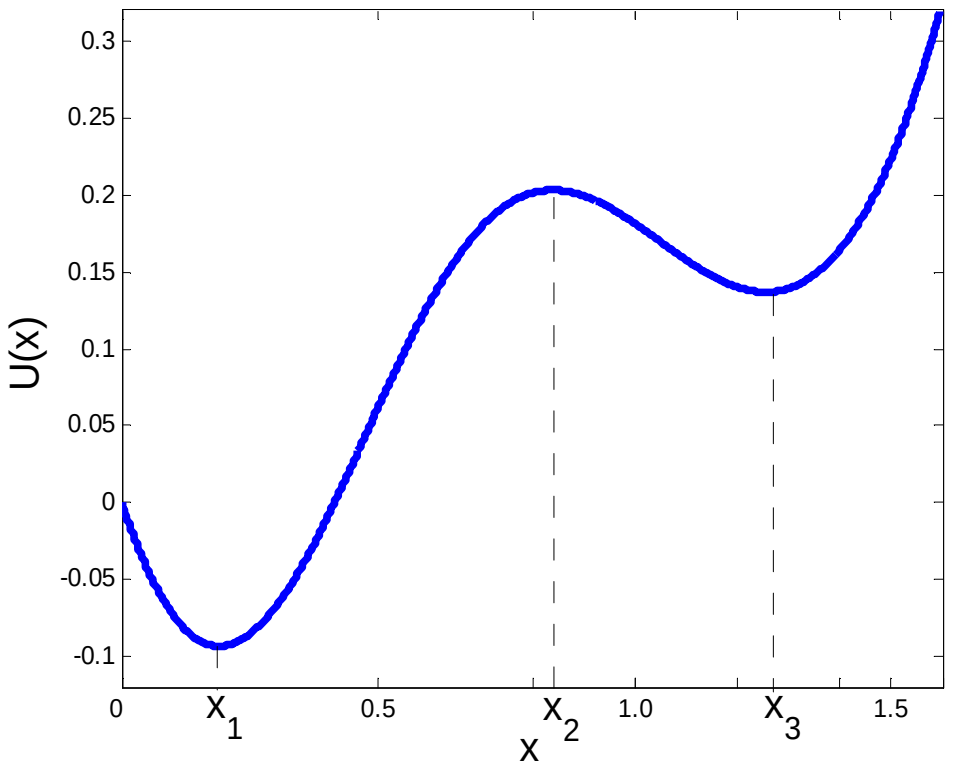

Fig.4. The potential function $U(x)$

Now consider non-Gaussian noise to the system (1), and then the stochastic version of gene regulatory model can be given as the Langevin differential equation:

$$\frac{dx}{dt}=\frac{k(2x^2+50x^4)}{25+29x^2+52x^4+4x^6}-rx+1+\frac{dL(t)}{dt}. \tag{2}$$

Here, $L(t)$ is the Lévy stable motion with stationary and independent increments on non-overlapping time intervals which can describe stochastic processes characterized by the occurrence of extremely long jumps, and its derivative denotes the Lévy noise. The process $L(t)$ is a generalized wiener process obeying Lévy distribution $L_{\alpha,\beta}(\zeta;D,\mu)$ which corresponds to a four-parametrical family of the probability density functions characterized by their Fourier transforms $\Phi(\theta)$, that is $\Phi(\theta)=F(L_{\alpha,\beta}(\zeta;D,\mu))=\int_{-\infty}^{+\infty}d\zeta e^{i\theta\zeta}L_{\alpha,\beta}(\zeta;D,\mu)$, then for $\alpha\in(0,1)\cup(1,2]$

$$\Phi(\theta)=\exp\left\{\left[-D^{\alpha}\left|\theta\right|^{\alpha}\left(1-i\beta sgn(\theta)\tan\frac{\pi\alpha}{2}\right)\right]+i\mu\theta\right\},$$

and for $\alpha=1$

$$\Phi(\theta)=\exp\left\{\left[-D\left|\theta\right|\left(1+i\beta sgn(\theta)\frac{2}{\pi}\ln\left|\theta\right|\right)+i\mu\theta\right]\right\}.$$

The parameter $\alpha(0<\alpha\le 2)$ denotes the asymptotic tail of the Lévy distribution and for $\alpha=2$ the Lévy stable distribution becomes a Gaussian distribution. $\mu\,(-\infty\le\mu\le\infty)$ is the center or location parameter which denotes the mean value of the distribution, and the mean of the distribution exists and equals $\mu$ as $1<\alpha\le 2$ [20]. The parameter $\beta\,(-1\le\beta\le 1)$ is the skewness parameter defining the degree of asymmetry of the distribution, namely, the Lévy distribution is symmetric for $\beta=0.0$ and asymmetric for $\beta\neq 0.0$ and $D$ represents the noise intensity [21]. The following context will be constrained to the case of $1<\alpha\le 2, \mu=0$ and $-1\le\beta\le 1$. The probability densities for different values of $\alpha$ and $\beta$ are shown in Fig.5 ($D=1.0, \mu=0$). In Fig.6, we present the sample trajectory of Gaussian process and Lévy process. Some visible jumps can be observed in Fig.6(b) so that Lévy noise is applicable to describe unpredictable jump changes of the random environment.

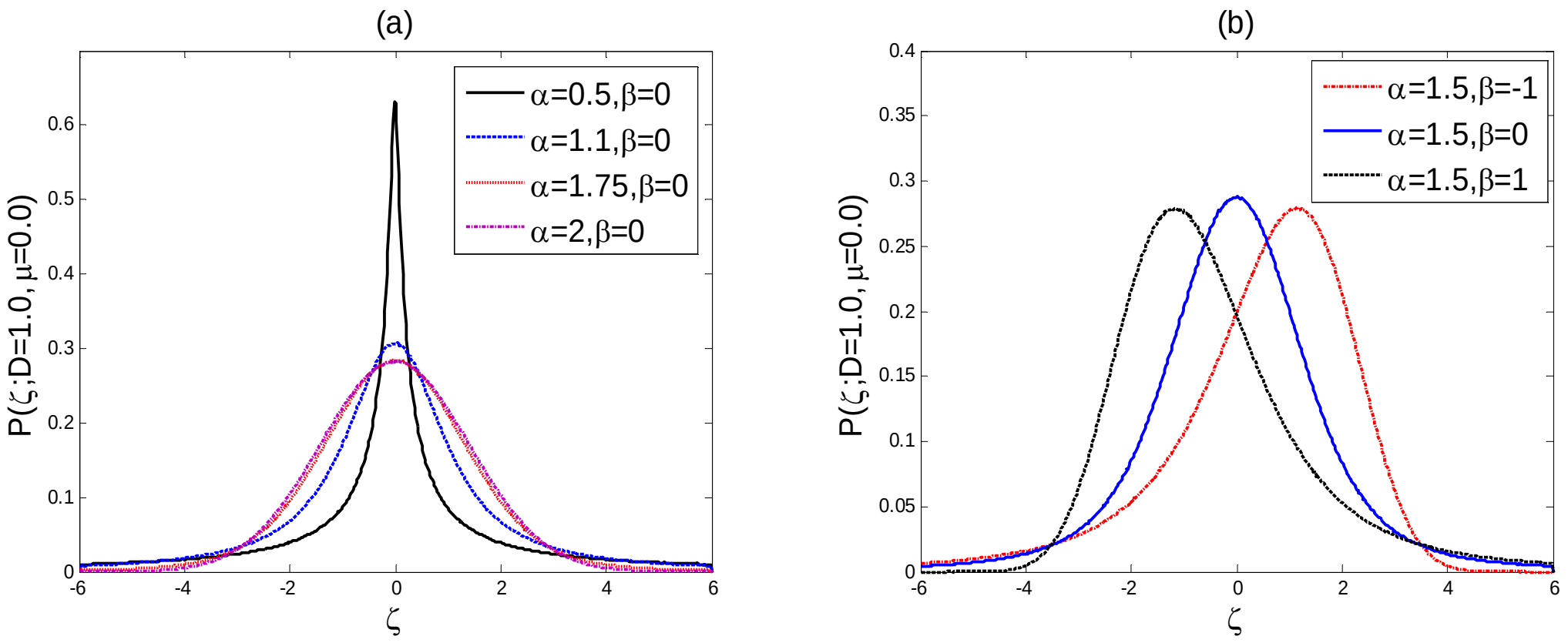

Fig.5. (a) Probability density functions of Lévy stable noise for different values of $\alpha$ ($\beta$=0,$D=1.0$, $\mu=0$); (b) Probability density functions of Lévy stable noise for different values of $\beta$ ($\alpha$=1.5,$D=1.0$, $\mu=0$).

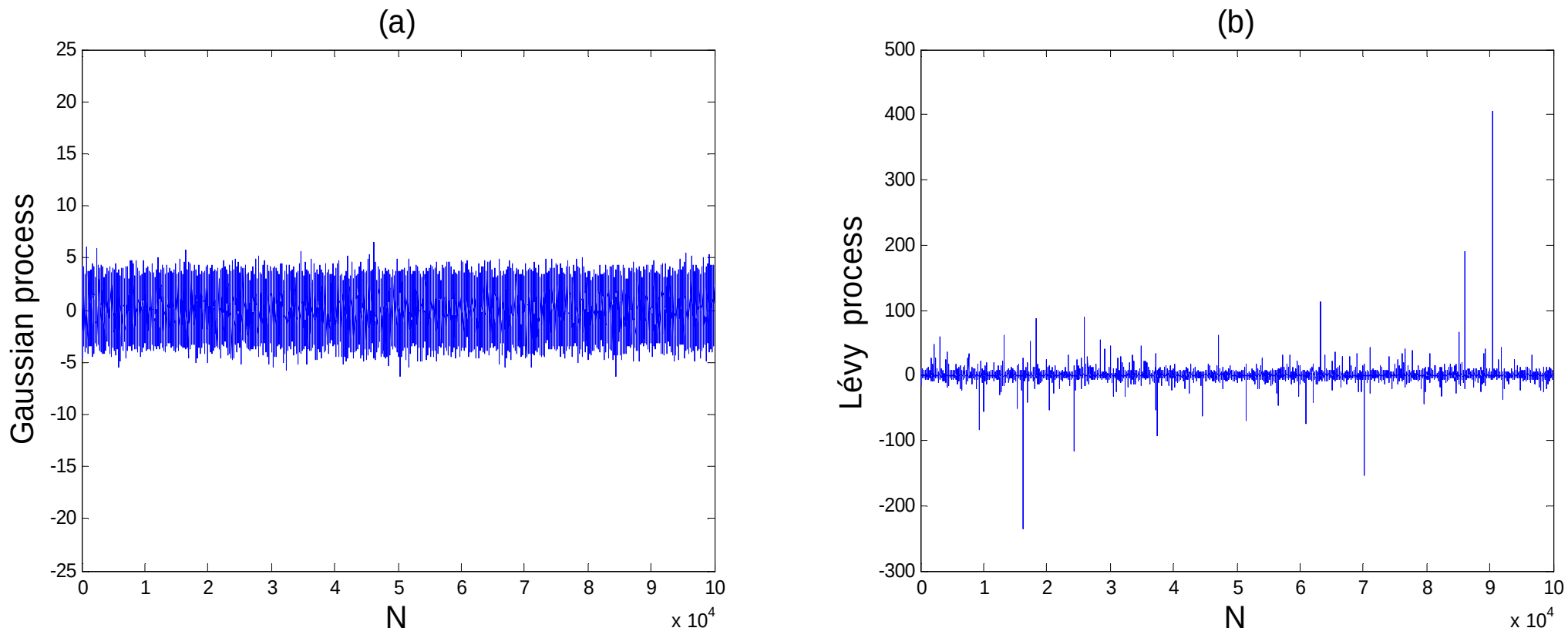

Fig.6. (a) Sample trajectory of Gaussian process ($\alpha$=2,$\beta$=0.2,$D=1.0$, $\mu=0$);(b) Sample trajectory of Lévy stable process($\alpha$=1.75,$\beta$=0.2,$D=1.0$, $\mu=0$).

The corresponding fractional Fokker-Plank equation of Eq.(2) is [22]:

$$\frac{\partial}{\partial t}P(x,t)=-\frac{\partial}{\partial x}F(x)P(x,t)+(\frac{1}{2}+\frac{\beta}{2})D(t)D_{a+}^{\alpha}P(x,t)+(\frac{1}{2}-\frac{\beta}{2})D(t)D_{b-}^{\alpha}P(x,t), \quad (3)$$

where $F(x)=\frac{k(2x^2+50x^4)}{25+29x^2+52x^4+4x^6}-rx$, $D(t)$ is the time-dependent coefficient of dispersion. $D_{a+}^{\alpha}$ and $D_{b-}^{\alpha}$ can be given as:

$$D_{a+}^{\alpha}P(x,t)=\frac{1}{\Gamma(2-\alpha)}\frac{\partial^2}{\partial x^2}\int_a^x\frac{P(\eta,t)d\eta}{(x-\eta)^{\alpha-1}},$$

and

$$D_{b-}^{\alpha}P(x,t)=\frac{1}{\Gamma(2-\alpha)}\frac{\partial^2}{\partial x^2}\int_x^b\frac{P(\eta,t)d\eta}{(x-\eta)^{\alpha-1}},$$

which stand for the left and right Riemann-Liouville space fractional derivatives of order $\alpha$ for probability density function $P(x,t)$, $x \in [a,b]$ with $a$ and $b$ the real numbers. In our paper, $x$ denotes concentration of $\lambda$-repressor and then the interval is constrained to $[0,+\infty]$.

$L(t)$ becomes a symmetric Lévy process with $\beta$=0 and Eq. (3) will be rewritten as:

$$\frac{\partial}{\partial t}P(x,t) = -\frac{\partial}{\partial x}F(x)P(x,t) + D(t)\frac{\partial^\alpha P(x,t)}{\partial |x|^\alpha}, \tag{4}$$

where $\partial^\alpha / \partial |x|^\alpha$ is the Riesz space-fractional derivative which can be defined as [23]:

$$\frac{\partial^\alpha P(x,t)}{\partial |x|^\alpha} = -\frac{D_{a+}^{\alpha} P(x,t) + D_{b-}^{\alpha} P(x,t)}{2\cos(\pi\alpha/2)}.$$

In the case of $\alpha = 2$, $L(t)$ reduces to the usual Brownian motion and Eq. (4) will reduce to the standard Fokker-Planck equation:

$$\frac{\partial}{\partial t}P(x,t) = -\frac{\partial}{\partial x}F(x)P(x,t) + D(t)\frac{\partial^2 P(x,t)}{\partial |x|^2}.$$

In Fig.7, we present the result of a numerical simulation in the case of Gaussian noise, and one can find, the switch starts at the ‘‘off’’ position by tuning the noise strength to a very low value. However, when we increase the noise intensity, the upper state becomes populated, corresponding to a concentration increase and a switch from the “off” to the “on” position and the result is fully consistent with the literature [7].

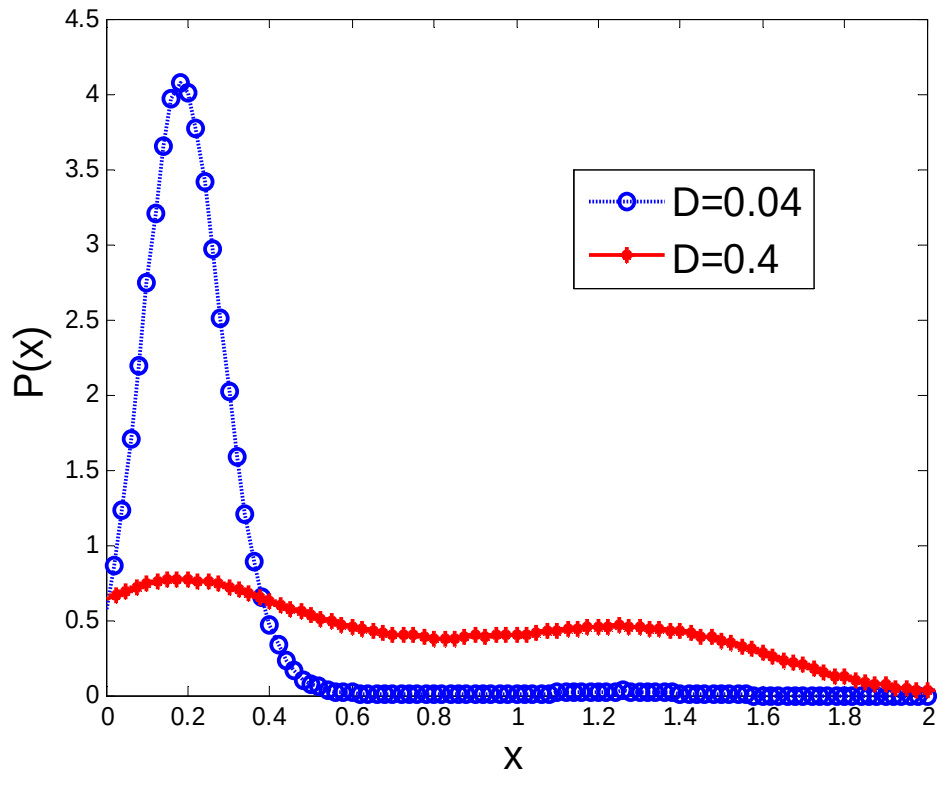

Fig.7 SPDs under Gaussian noise for noise strengths $D = 0.04$ and $D = 0.4$

However, in the case $1 < \alpha < 2$ the situation is radically different. The sample curves of the process are no longer continuous and the extreme jumps may contribute to the switch between two states. It’s very difficult to get the analytical solutions of Eq.(3), therefore, we resort to Monte-Carlo simulations where SPDs are extracted from ensembles of $N = 10^6$ trajectories of a given length $t$. The value of $t$ is chosen by trial and comparison of numerical estimations of the time dependent probability distribution function $P(x,t)$ for various $t$, as shown in Fig.8, we can find that all the systems preserve the states after the time $t = 10s$. Therefore, it can be regarded as stationary, i.e. $P_{st}(x) = P(x,10)$. Then, the time course of CI concentration and the SPDs are plotted by simulating different parameters of Lévy noise to observe properties of the switch between stable states.

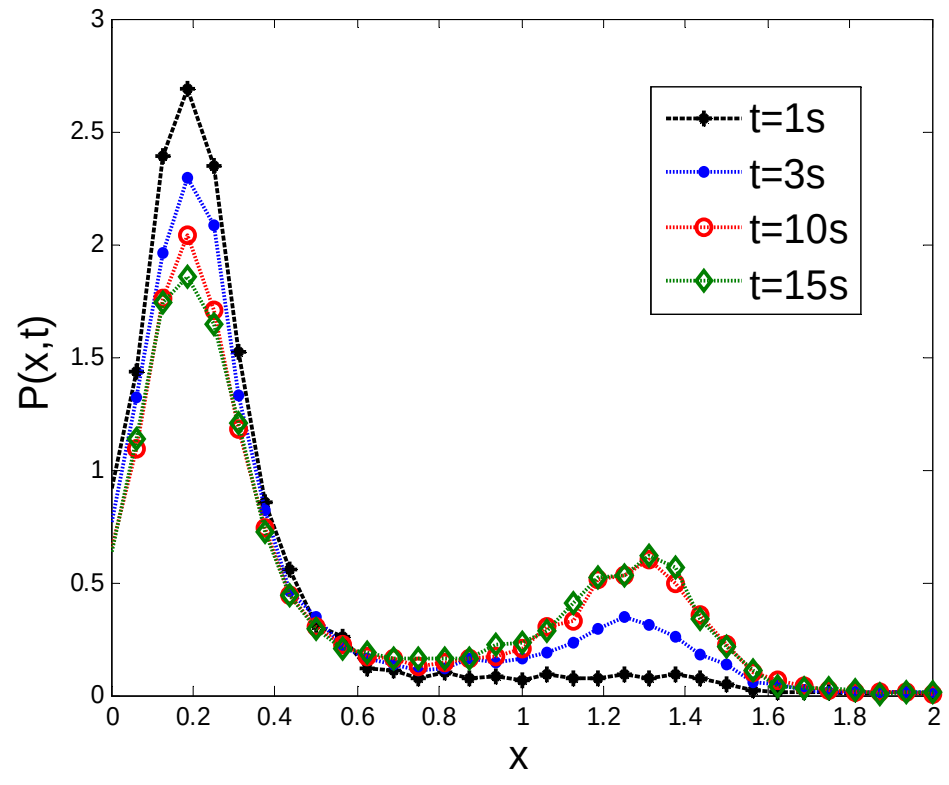

Fig.8 $P(x,t)$ at different times for $\alpha = 1.5, D = 0.2, \beta = -0.3$.

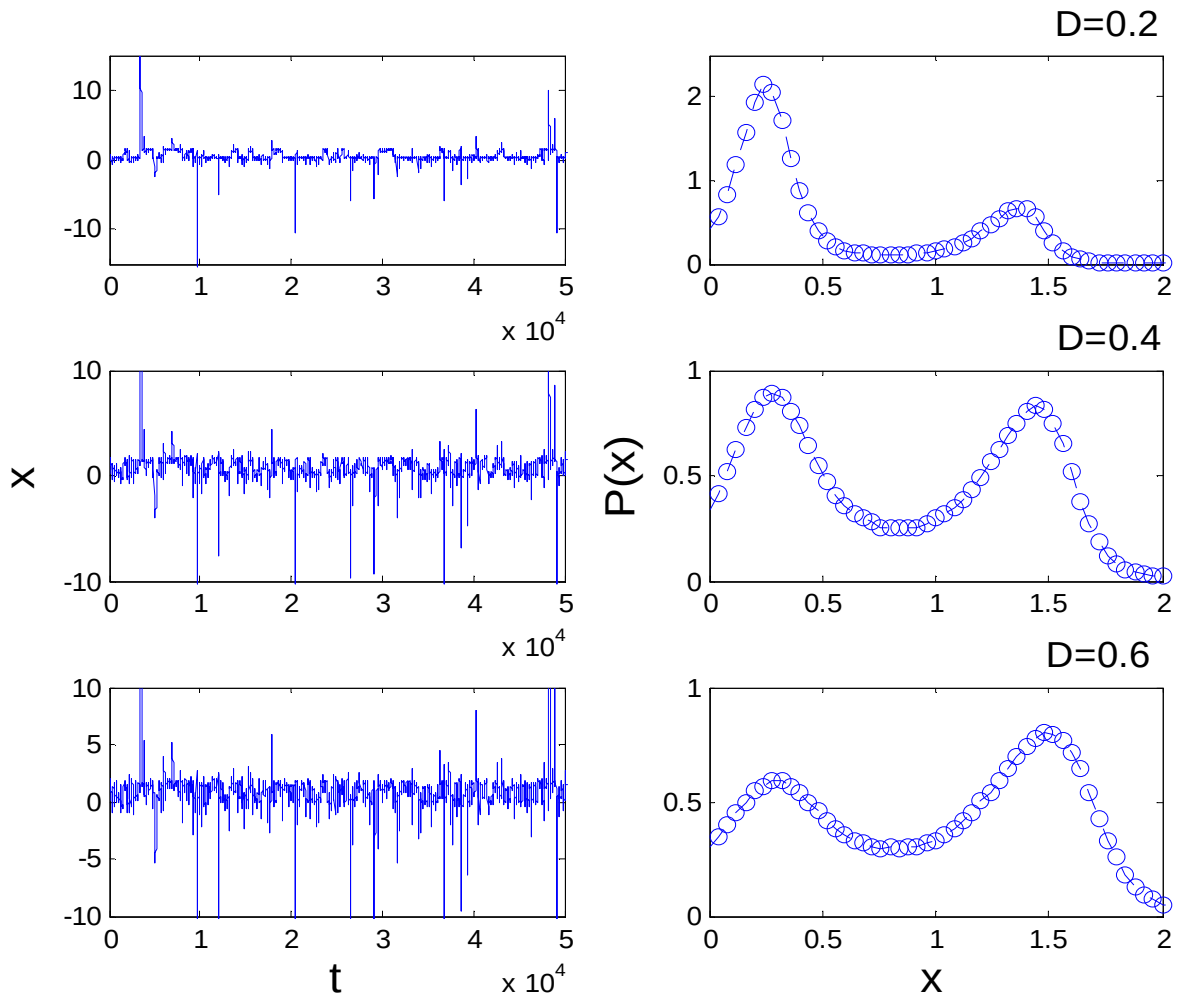

Fig 9. Sample paths and SPDs for different noise intensities.

From top to bottom $D = 0.2, D = 0.4$ and $D = 0.6$ ( $\alpha = 1.4, \beta = -0.6$ ).

In Fig. 9, the time course of CI concentration and the SPDs are plotted for different noise intensities respectively. It is shown that when the noise intensity $D$ is small, the peak at the lower steady state is much higher, indicating the CI concentration concentrates on the low concentration state, i.e. the switch is “off”. Increasing the noise intensity, the peak on the side of the high values of $x$ becomes higher while the lower peak diminishes in height. Until certain noise intensity, the high concentration state becomes more populated and we can say the “off” to “on” switching occurs, which demonstrates the noise intensity can induce positive feedback to increase CI concentration. Moreover, we present the average CI concentration $\langle x \rangle$ as a function of the noise intensity as Fig.10, in which we can find the effect of noise intensity directly. Here, $\langle x \rangle$ is defined as $\langle x \rangle = \int_0^{+\infty} x P_{st}(x) dx$. It can be seen that increasing $D$ means growing of the CI concentration. All facts reveal that noise intensity leads to a high level of expression and so it can be regarded as a control parameter of the gene switch.

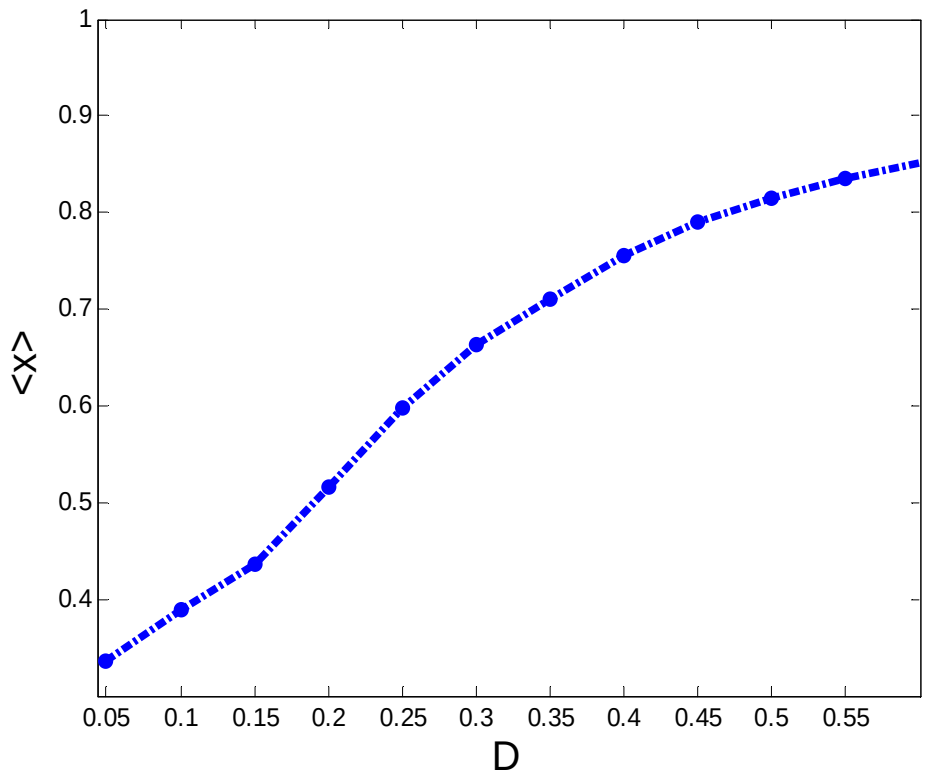

Fig.10 The average CI concentration versus $D$

The SPDs versus stability index are shown in Fig. 11. Initially, there is a visible peak in the higher state and the other is negligible indicating a correspondingly high value of the CI concentration, that is to say, the switch is in the ‘‘on’’ position at lower value of $\alpha$. With the increase of stability index, the peak at the low value of $x$ grows in height and finally prevails, corresponding to a flipping of the switch from “on” to “off ” position and leading a CI concentration decrease. The results indicate that, when the other parameters are fixed, increasing the stability index can induce a switch. As different kinds of noises can be represented by different values of stability index, the results obtained also provide compelling evidence that there is a noticeable difference in dynamics under Gaussian and non-Gaussian Lévy noise. Fig.12 presents the average CI concentration versus stability index. It clearly indicates that, with the increase of $\alpha$, the average concentration of CI gradually moves to the lower value which means a switch from “on” to “off”. The conclusion is consistent with the analyses of SPDs.

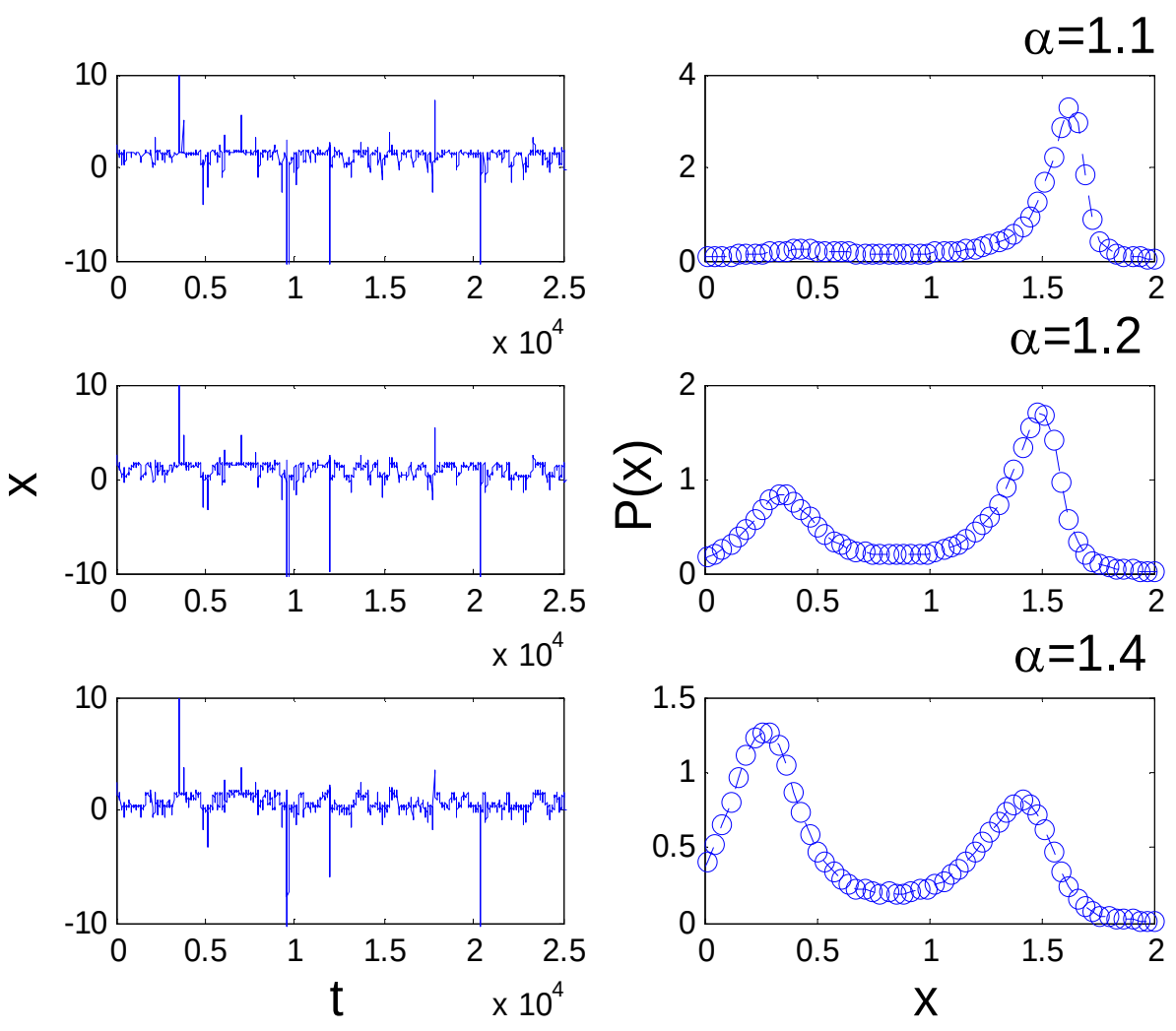

Fig.11 Sample paths and SPDs versus stability index.

From top to bottom $\alpha = 1.1, \alpha = 1.2$ and $\alpha = 1.4 (D = 0.3, \beta = -0.6)$.

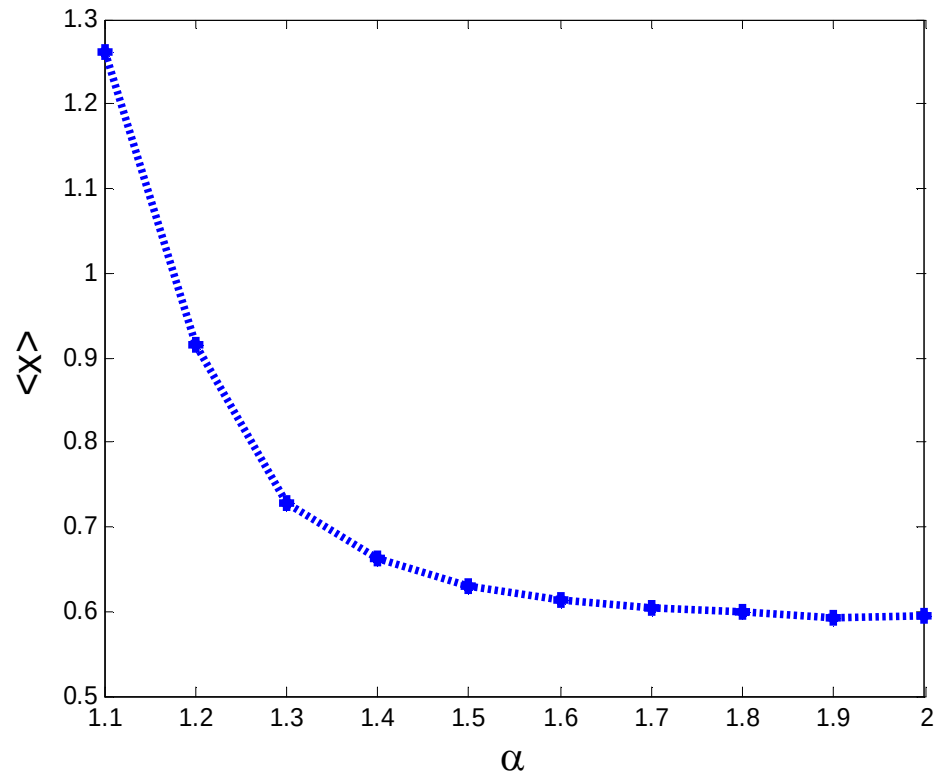

Fig. 12 The average CI concentration versus $\alpha$

Fig.13 depicts the time evolution of the CI concentration and the SPDs for three different values of the skewness parameter $\beta = -0.8, \beta = -0.3$ and $\beta = 0.2$. At the lowest $\beta$ value, the peak on the right is much higher, which means that the switch is in the ‘‘on’’ position indicating a correspondingly high value of the concentration. However, we note that increasing the skewness parameter causes the low concentration state to become populated. It reveals that the switch turns from “on” to “off” which leads to a decrease of the CI concentration. Fig.14 shows the average CI concentration versus $\beta$ and the result is in accordance with the analyses above. Therefore, we conclude that the skewness parameter can induce a switch process resulting in a decline of the CI concentration in the Lévy case. However, as $\alpha$ approaches 2, the skewness parameter loses its effect and becomes irrelevant in the Gaussian case as Fig. 15 shows.

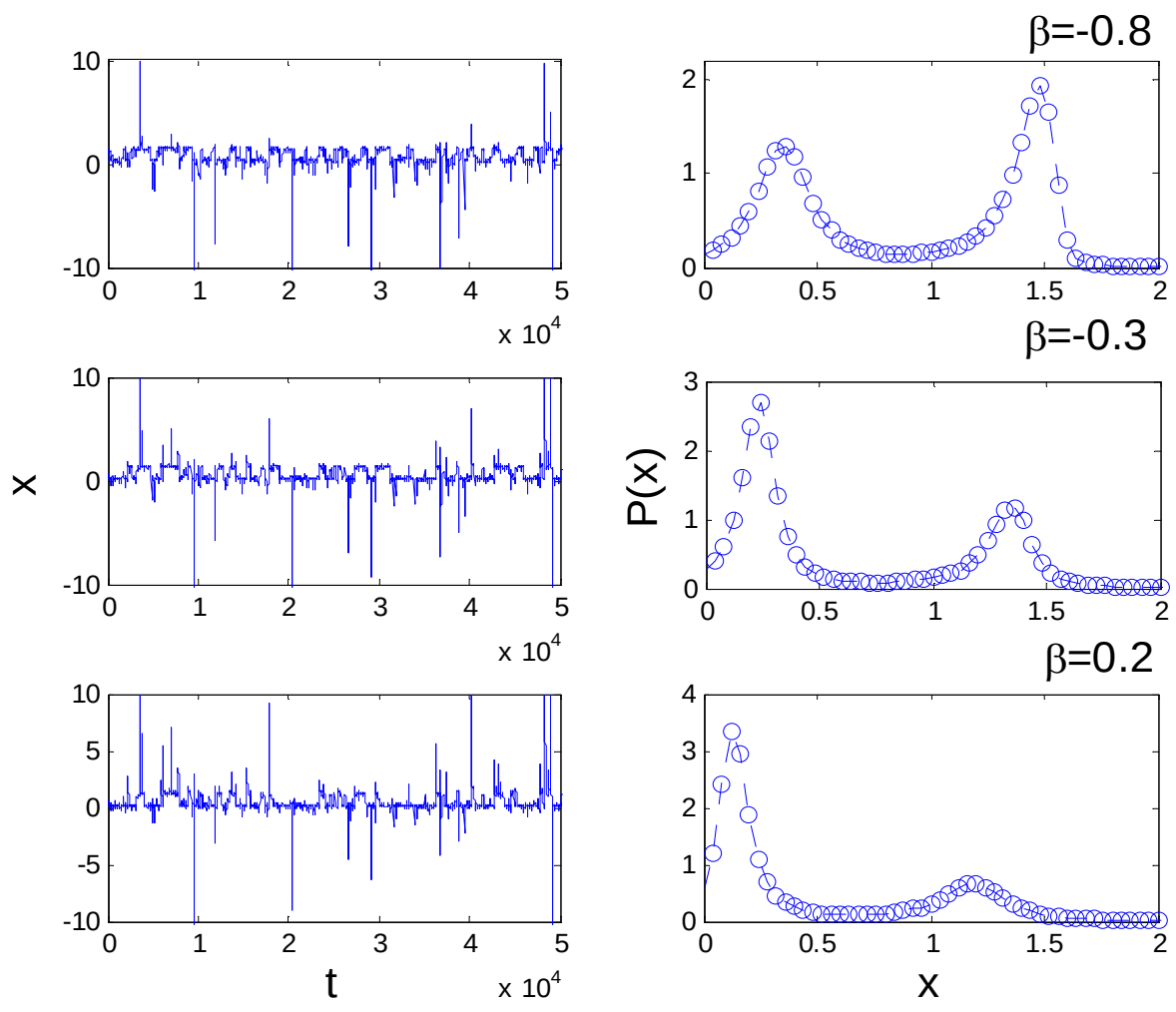

Fig.13 The time evolution of the CI concentration and the SPDs at

$\beta = -0.8, \beta = -0.3$ and $\beta = 0.2$ ( $D = 0.2, \alpha = 1.2$ ).

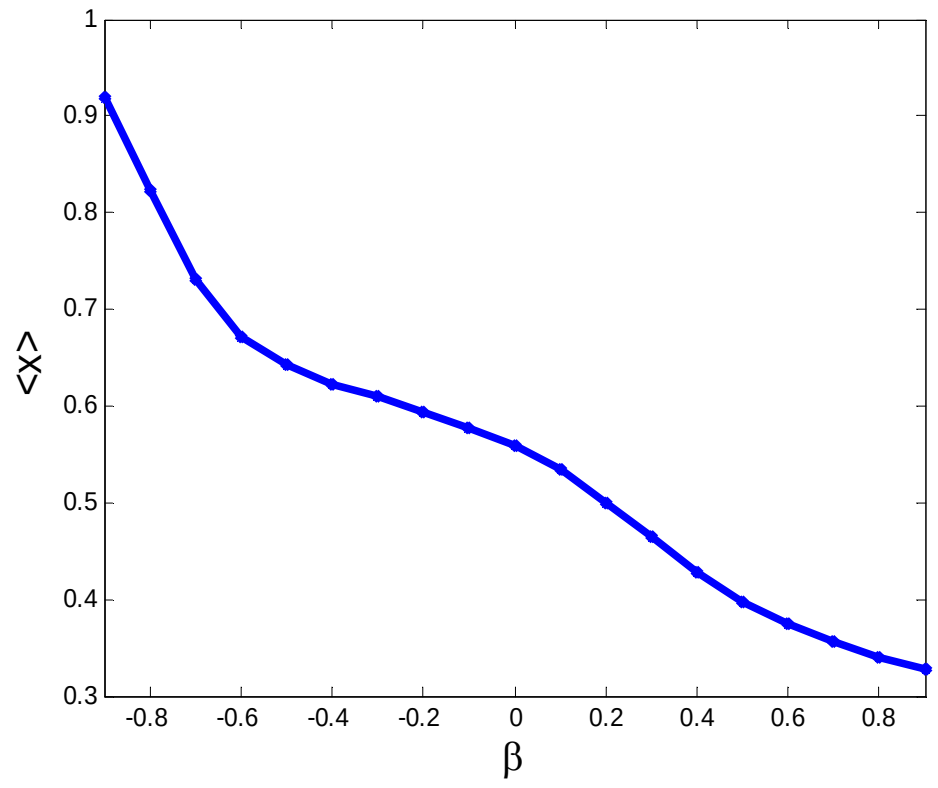

Fig.14 The average CI concentration versus $\beta$

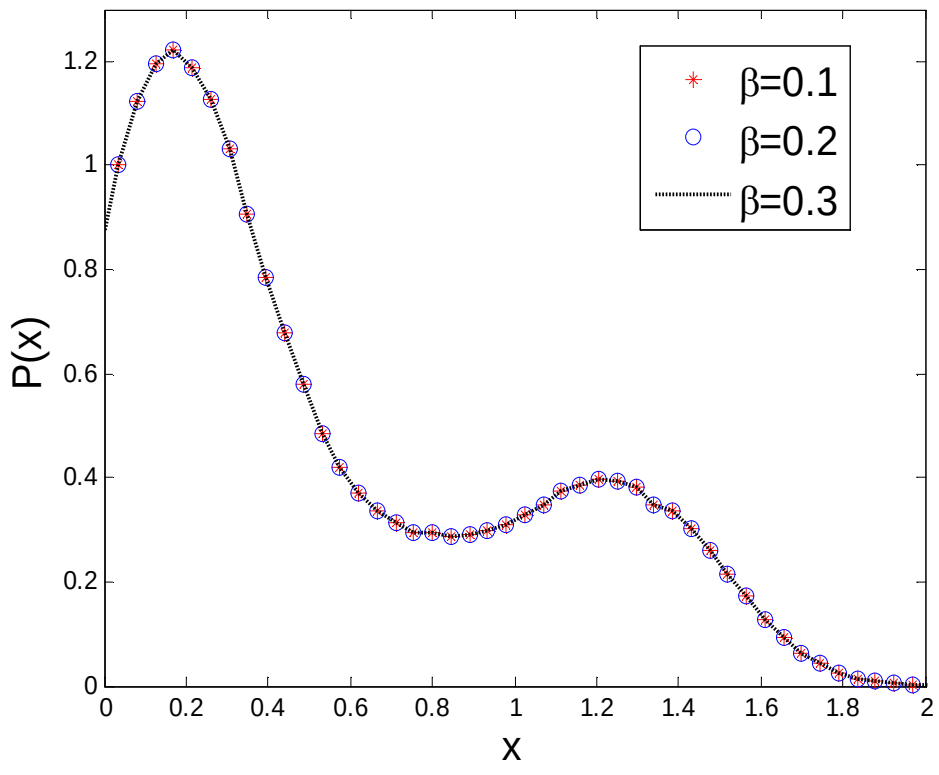

Fig. 15 SPDs versus $\beta$ in the Gaussian case ( $D = 0.2$ )

# IV. Mean First Passage Time

In this section, we are interested in the time between switches from low to high concentration and vice versa. This time, which is referred to as the MFPT, is a random variable defined as the time elapsed until a random process reaches a prescribed state for the first time. Now we study the influences of Lévy noise on the MFPT in two directions.

First, we compute the MFPT from $x_1$ to $x_3$ to consider the influences of Lévy noise on the time that is used to switch from low to high concentration. The event is equivalent to the increase of CI concentration. By solving the equation $U'(x)=0$, we obtain the two stable states of the system: $x_1 = 0.1937, x_3 = 1.2587$. In the simulation, we let the starting point at the lower state, and then calculate the average time over 10000 times for the concentration to reach $x_3$. The MFPT versus noise intensity, stability index and skewness parameter is shown in Fig.16.

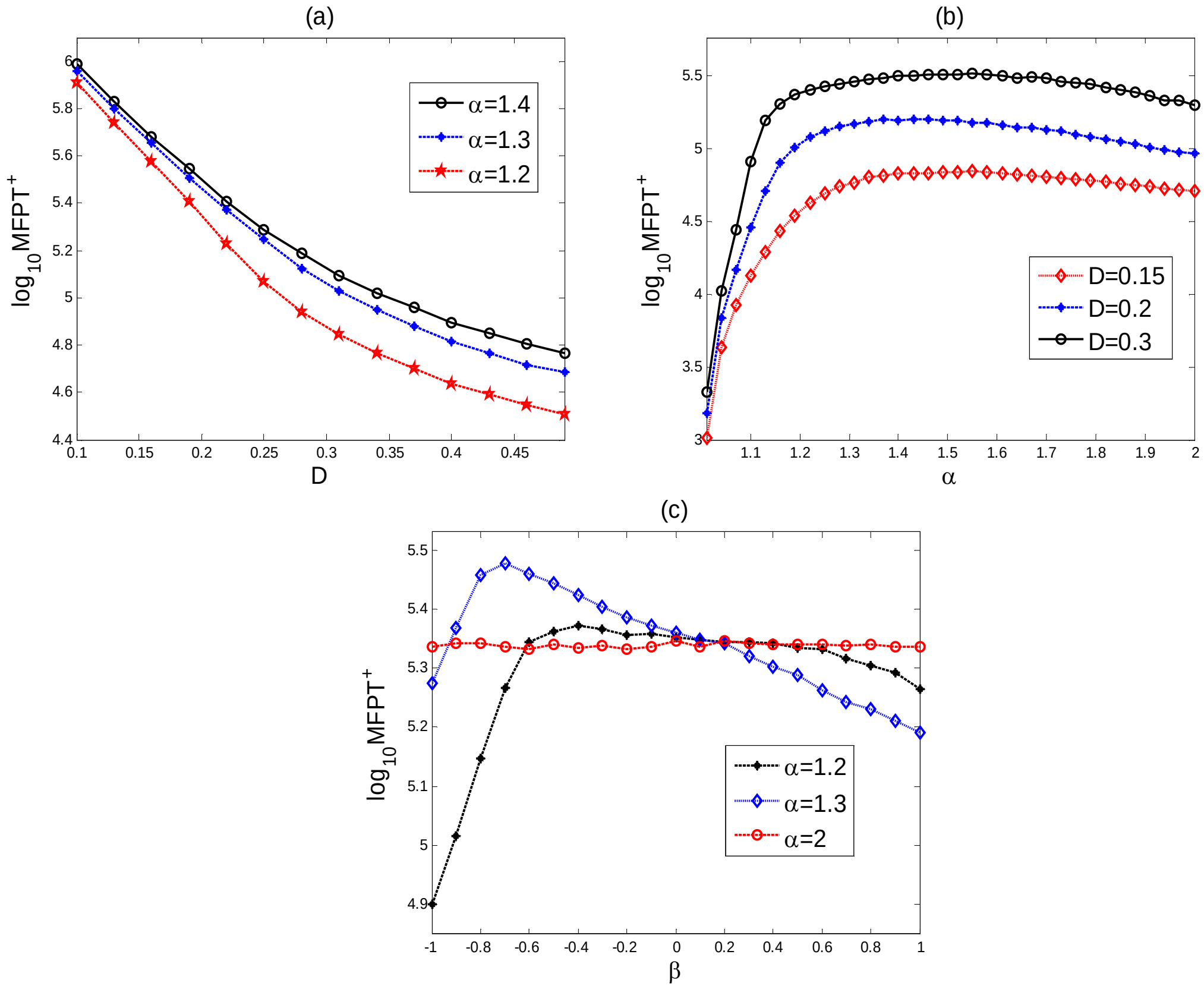

Fig.16 (a) MFPT $T_+(x_1 \to x_3)$ versus noise intensity

(b) MFPT $T_+(x_1 \to x_3)$ versus stability index

(c) MFPT $T_+(x_1 \to x_3)$ versus skewness parameter

In Fig.16(a), it can be seen that the average time decreases due to the increases of noise intensity. With regard to biological aspects, when the CI concentration is initially at the low value state, the transition from the low concentration state to the high concentration state would become much easier with the increase of noise intensity. That means the noise intensity can speed up the expression of CI.

Fig.16(b) shows that as $\alpha$ increases, a critical stability index exists at which the system (5) needs the longest time to transit. On the left of this critical value, the MFPT is increasing while in another side it is decreasing, however, the rate of increase is much larger than the decrease rate. Therefore, increasing in stability index first makes it more favorable for the CI expression, when the parameter crosses a certain threshold value, it will hamper the expression of CI.

Similarly, it is shown in Fig.16(c) that an increasing $\beta$ leads to a higher value of time spent at first. However, when the MFPT reaches a maximum, it will decreases with the increase of $\beta$ corresponding to a shorter time spent. Therefore, it can be concluded that, the transition from the low concentration state to the high concentration one which acts as gene expression is repressed when considering small skewness parameters. After reaching a certain value, the skewness parameter begins to play a positive role in switching from "off" to "on" to enhance the gene expression. From Fig.15(c), we can also reconfirm that the skewness parameter has no effect on the system (5) in the Gaussian case ($\alpha = 2$).

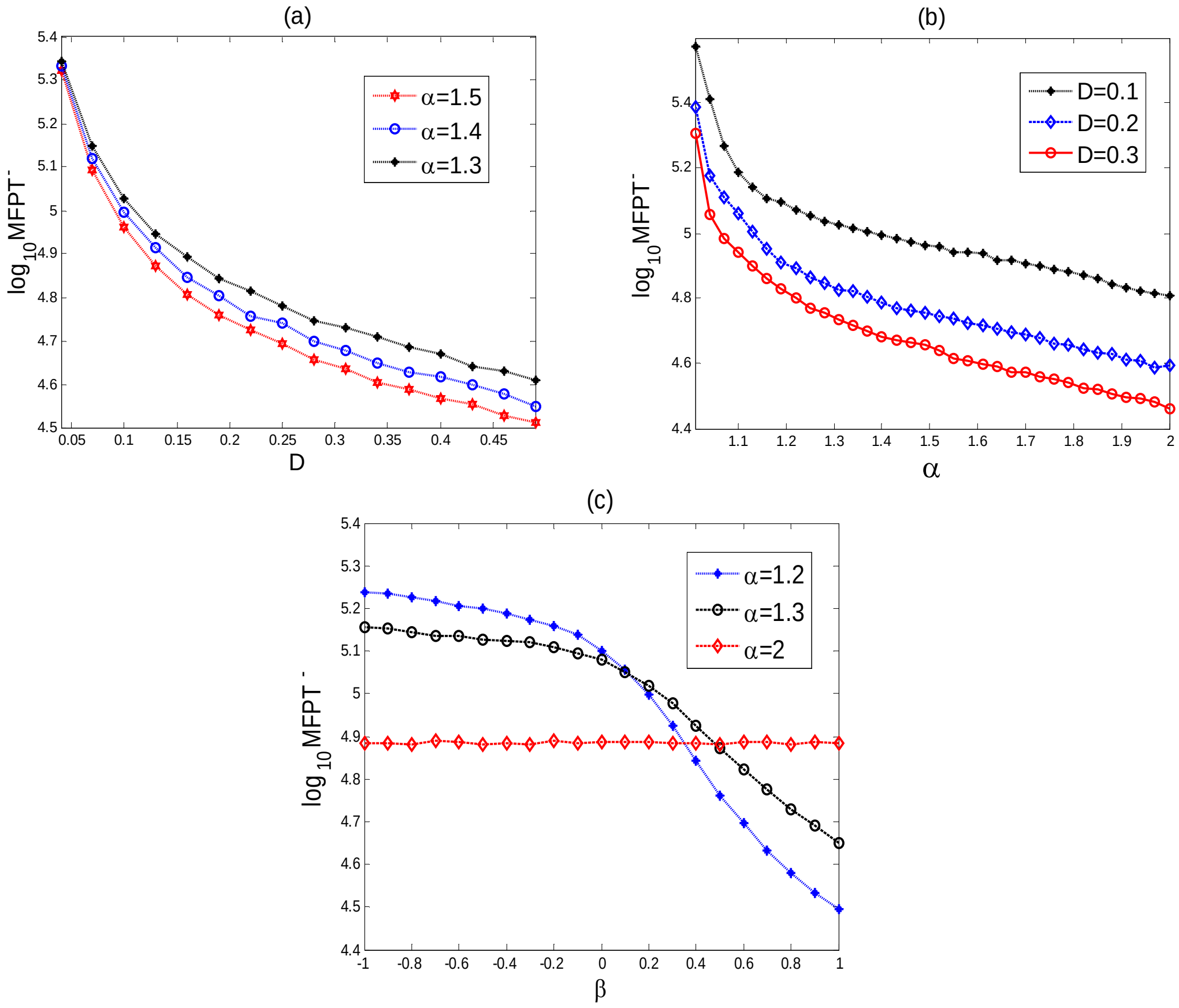

Fig.17 (a) MFPT $T_{_}(x_3 \to x_1)$ versus noise intensity

(b) MFPT $T_{-}(x_3 \to x_1)$ versus stability index

(c) MFPT $T_{-}(x_3 \to x_1)$ versus skewness parameter

To evaluate how Lévy noise affects the degradation time of CI, the MFPT is calculated by initializing the system at the high concentration state and calculating the average time it takes to reach the low concentration one. Fig.17(a) shows that the MFPT decreases with the increase of the noise intensity. Therefore, with an initial position at the high concentration, the MFPT for the system (5) to reach the low concentration decreases due to the noise intensity. In biology, we can infer that the degradation events will occur at earlier times for larger fluctuations.

In Fig.17(b), we can see that with an initial position at the high concentration, the MFPT for the system to reach the low concentration decreases by increasing stability index. Therefore for larger stability index the repressor CI is more obvious degradation. In summary, if the concentration of CI is higher at first, Lévy noise with larger stability index will play a more positive role in the degradation process to reduce the CI concentration.

Fig.17(c) presents the MFPT versus the skewness parameter. The behavior revealed from the result can be summarized as follows. The MFPT value is the largest at $\beta = -1$ and decreases as $\beta$ increases. This implies that when the system starts at the upper state with a correspondingly high value of the concentration and its transition to the lower state represents a low value of the concentration is facilitated by the skewness parameter. The larger value of β, the easier to switch form the high expression state to low expression one corresponding to the degradation of CI. When we consider the Gaussian distribution corresponding to the Lévy stable one with $\alpha = 2$, it also shows that the skewness parameter is irrelevant in this case.

# V. Concluding Remarks

Random fluctuations associated with a dynamic system are often assumed to be Gaussian noise in nature. However, Gaussian noise is unable to describe the natural fluctuations that characterized by the occurrence of extremely long jumps. Thus, it is important to incorporate the Lévy noise into the study of stochastic fluctuations. In the paper, based on the kinetic model of genetic regulation system proposed by Hasty, the effects of fluctuations which can be regarded as Lévy noise in the degradation rate have been investigated through numerical computation. It is straightforward to see that the noise-induced switch is different for $\alpha$-stable Lévy noise from the usual Gaussian case. Not only noise intensity, but also stability index and skewness parameter play important roles in the genetic regulatory switch processes. For smaller value of noise intensity, there is a high population in the lower state and a correspondingly low value of the concentration indicates the switch is in the “off” position. Increasing the noise intensity causes the upper state to become populated, corresponding to a concentration increase and a flipping of the switch to the “on” position. For stability index, the switch starts at the “on” position and with the increase of stability index, $x$ is shifted from the high concentration state to the low concentration one, denoting a successful switch to the “off” position. Finally, we consider the effect of the skewness parameter and the switch process can also be induced from “off” to “on”. All these parameters can be regarded as control parameters in gene switches.

Further we have quantified the effects of Lévy noise on the switch time between the two stable steady states, which is often referred to as the MFPT. The MFPT of the process $x(t)$ to reach the high concentration state $x_3$ with initial condition $x(t=0)=x_1$ (the low concentration state) is studied as functions of noise intensity, stability index and skewness parameter. As indicated by the numerical results, the transition from the low concentration state to the high concentration one is enhanced by noise intensity which means $D$ can promote the expression of CI. Besides, one observes that with the increase of stability index and skewness parameter, the MFPT increases to maximum then followed by a decrease. The resonance-like behaviors show that an appropriate set of the two parameters can enhance the gene expression, whereas in the other set, they can repress the gene expression.

Then, we have examined how the parameters of Lévy noise affect the degradation time of CI. We start with the trajectory of CI concentration in the higher state indicating a high level of CI concentration. Thanks to Lévy noise, which drives the system towards the lower state, a transition leading to a degradation of CI concentration occurs. The average time for the transition was explored versus the parameters of Lévy noise. Results showed that MFPT in this direction decreases with the increase of noise intensity, stability index and skewness parameter. That is to say, all the parameters of Lévy noise play positive roles in the degradation process by reducing the degradation time. However, in the Gaussian case, the skewness parameter becomes irrelevant and has no impact on the MFPT.

In summary, we have investigated a gene transcriptional regulatory system subjected to non-Gaussian Lévy noise which is more realistic. The fluctuation induced switch processes, one of the most important properties of the stochasticity in the gene transcriptional regulatory system, have been discussed via considering SPDs of the CI concentration. We have also computed the switching time as a function of different parameters of Lévy noise to characterize the noise effects. The results clearly indicate that there is a noticeable difference between the influences of Lévy and Gaussian noises.

## Acknowledgments

This work was supported by the NSF of China (Grant Nos. 10972181, 11102157), Program for New Century Excellent Talents in University, the Shaanxi Project for Young New Star in Science & Technology and NPU Foundation for Fundamental Research and New Faculty and Research Area Project. We thank Barbara Gentz, Ilya Pavlyukevich and Jinqiao Duan for their valuable suggestions and discussions.

## References

[1] P Smolen, D Baxter, J Byrne, Mathematical modeling of gene networks, Neuron 26, 567–580(2000).

[2] J Hasty, D McMillen, F Isaacs, J Collins, Computational studies of gene regulatory networks: in numero molecular biology, Nature Reviews Genetics 2, 268-279(2001).

[3] N Kaminski, Z Bar-joseph, A Patient-Gene Model for Temporal Expression Profiles in Clinical Studies, Journal of Computational Biology 14(3), 324-338(2007).

[4] T Bose, S Trimper, Stochastic model for tumor growth with immunization, Phys. Rev. E 79, 051903(2009).

[5] H Hasegawa, Stochastic bifurcation in FitzHugh–Nagumo ensembles subjected to additive and/or multiplicative noises, Phys. D 237, 137–155(2008).

[6] P Hänggi, Stochastic resonance in biology: How noise can enhance detection of weak signals and help improve biological information processing, Chemphyschem 3, 285–290(2002).

[7] J Hasty, J Pradines, M Dolnik, J Collins, Noise-based switches and amplifiers for gene expression, Proceedings of the National Academy of Sciences 97, 2075-2080(2000).

[8] P Smolen, D Baxter, J Byrne, Frequency selectivity, multistability, and oscillations emerge from models of genetic regulatory systems, Am J Physiol Cell Physiol 274, C531-C542(1998).

[9] Q Liu, Y Jia, Fluctuations-induced switch in the gene transcriptional regulatory system, Phys. Rev. E 70, 041907 (2004)

[10] M Huang, J Wu, Y Luo, K Petrosyan, Fluctuations in gene regulatory networks as Gaussian colored noise, JCP: BioChemical Physics 4, 04B606 (2010).

[11] A Edwards, et al., Revisiting Lévy flight search patterns of wandering albatrosses, bumblebees and deer, Nature 449, 1044–1048(2007).

[12] F Bartumeus, et.al, Animal search strategies: a quantitative random-walk analysis, Ecology 86(11), 3078–3087(2005).

[13] M Lomholt, T Ambjörnsson, R Metzler, Optimal target search on a fast folding polymer chain with volume exchange, Phys. Rev. Lett 95, 260603(2005).

[14] M Sokolov, R Metzler, K Pant, M Williams, Target search of N sliding proteins on a DNA, Biophysical Journal 89, 895–902(2005).

[15] T Schötz, R Neher, U Gerland, Target search on a dynamic DNA molecule, Phys. Rev. E 84, 051911(2011).

[16] M Assaf, E Roberts, Z Luthey-Schulten, Determining the stability of genetic switches: explicitly accounting for mRNA noise, Phys. Rev. Lett 106, 248102(2011).

[17] H Risken, The Fokker–Planck Equation, Springer-Verlag, Berlin, 1996.

[18] T Kepler, T Elston, Stochasticity in Transcriptional Regulation: Origins, consequences and mathematical representations, Biophysical Journal 81, 3116–3136(2001).

[19] A Walczak, M Sasai, P Wolynes, Self-consistent Proteomic Field Theory of Stochastic Gene Switches, Biophysical Journal 88, 828–850(2005).

[20] R Weron, Lévy-stable distributions revisited: tail index > 2 does not exclude the Lévy-stable regime, International Journal of Modern Physics C 12(2), 209-223(2001).

[21] A Janicki, A Weron, Simulation and chaotic behavior of $\alpha$-stable stochastic

processes, Marcel Dekker, New York, 1994.

[22] D Benson, S Wheatcraft, M Meerschaert, The fractional-order governing equation of Lévy motion, Water Resources Research 36, 1413-1424(2000).

[23] S Jespersen, Lévy fights in external force fields: Langevin and fractional Fokker-Planck equations and their solutions, Phys. Rev. E 59, 2736–2745(1999).